%% file: 00-paper.tex
\documentclass[manuscript, nonacm]{acmart}

\input{z-commands}

\usepackage{calc, enumitem} 
\usepackage{soul}
\usepackage{xcolor, colortbl}
\usepackage{multirow}
\usepackage{rotating}
\usepackage{subcaption}
\usepackage{stfloats}
\usepackage{wasysym} 
\usepackage{totcount}\regtotcounter{fixmeTotalCounter}
\usepackage{datetime2} 

\copyrightyear{2023} 
\acmYear{2023} 

\begin{document}

\input{000-titleAuthorAbstract.tex}


%

\maketitle

\input{documentBody/01-Introduction}

\input{documentBody/02-RelatedWorks}
\input{documentBody/03-Method}
\input{documentBody/04-ResultsDiscussion}
\input{documentBody/05-Conclusion}

\bibliographystyle{ACM-Reference-Format}
\bibliography{00-paper,Biblio-Database,sample}

\newpage
\pagestyle{empty}
\end{document}

%% file: z-commands.tex
\newcommand{\STT}[1]{{%
\leavevmode\color{black}
\textsl{STT}}} 
\newcommand{\STTlong}[1]{{%
\leavevmode\color{black}
Scores Through-Time}}

\newcommand{\REDACT}[1]{$\Box\Box\Box$} 

\newcommand{\redactCollege}[1]{[a U.S. University]}  

\newcounter{boldifyCounter}
\newcounter{fixmeSectionCounter}
\newcounter{fixmeTotalCounter}
\makeatletter
\@addtoreset{fixmeSectionCounter}{section}
\@addtoreset{fixmeSectionCounter}{subsection}
\@addtoreset{boldifyCounter}{section}
\@addtoreset{boldifyCounter}{subsection}
\makeatother

\newcommand{\boldify}[1]{}
\ifdefined\boldifyON
	\renewcommand{\boldify}[1]{
        \par\noindent
		\stepcounter{boldifyCounter}
		\textbf{{\color{green}**}
		~\arabic{section}.\arabic{subsection}.\arabic{boldifyCounter}
		: #1} 
	}
\fi 

\newcommand{\reportOnFIXME}{%
    \newcount\iterCounter
    \iterCounter=1
    \newcount\endCounter
    \endCounter=\totvalue{fixmeTotalCounter}
    \advance \endCounter +1
    There are 
    {\color{red}\total{fixmeTotalCounter}} 
    FIX\_ME\\
    links:
    \loop
        \hyperlink{fixTag\the\iterCounter}{\#\the\iterCounter}
        \advance \iterCounter +1
    \ifnum \iterCounter < \endCounter
    \repeat
}

\newcommand{\FIXME}[1]{} 
\ifdefined\fixmeON
	\renewcommand{\FIXME}[1]{\par\noindent
		\stepcounter{fixmeSectionCounter}\stepcounter{fixmeTotalCounter}
		{\color{red}\fbox{\color{black}
			\parbox{.99\linewidth}{
				\textbf{\hypertarget{fixTag\thefixmeTotalCounter}{FIXME}	\arabic{section}.\arabic{subsection}.
        		\arabic{fixmeSectionCounter} (\color{red}
        		\#\arabic{fixmeTotalCounter}):} #1}}
        }
	}
\fi 

\newcommand{\FIXED}[1]{}
\ifdefined\fixmeON
	\renewcommand{\FIXED}[1]{\par\noindent%
		{\color{black}\fbox{\color{black}%
			\parbox{.99\columnwidth}{%
				\color{blue}#1}}%
        }
	}
\fi 

\newcommand{\draftStatus}[1]{}
\ifdefined\draftStatusON
	\renewcommand{\draftStatus}[1]{
        \hfill **#1
	}
\fi 

%% file: 000-titleAuthorAbstract.tex
\title{Understanding Driver Agency in RideSharing}

\author{Iyadunni Adenuga}
\email{ija5027@psu.edu}
\affiliation{%
  \institution{%
  Pennsylvania State University}
  \streetaddress{Westgate Building}
  \city{University Park}
  \state{PA}
  \country{USA}
  \postcode{16802}
}

\author{Benjamin Hanrahan}
\email{bvh10@psu.edu}
\affiliation{%
  \institution{%
  Pennsylvania State University}
  \streetaddress{Westgate Building}
  \city{University Park}
  \state{PA}
  \country{USA}
  \postcode{16802}
}

\renewcommand{\shortauthors}{I. Adenuga}
\renewcommand{\shorttitle}{Driver Agency}

\begin{abstract}
Agency is an important human characteristic that users of automated complex technologies are usually denied.
This affects the user's experience leading to decreased satisfaction and productivity.
In this paper, we consider the ridesharing context and interviewed 7 drivers to understand the controls that would improve the agency they feel.
The results show that they desire transparency, community and an effective ability to seek redress.
\end{abstract}

%% file: documentBody/01-Introduction.tex
\section{Introduction}

Many complex systems affect our lives directly or indirectly, while running automated algorithms that are not created to be explainable, answerable or amenable to people. 
These complex systems are usually, not adaptable and are not designed to consider or react to possible imprecise actions from users \cite{hoddinghaus2021automation}. 
Today, progress in the development of complex systems, such as systems that use artificial intelligence (AI) technology, is accompanied by the automation and  erasure of human inputs and efforts. 
For laborious, manual, repetitive tasks, automation is usually welcome but people notice and care about the loss of agency \cite{huh2010incorporating}. 
End-users are encountering these ever-increasing number of complex algorithms in their daily lives in a wide range of applications and so, there have been many calls for \textit{algorithmic transparency} in a number of domains \cite{oduor2008effects, 10.1145/2858036.2858402, 10.1145/3290605.3300724}. 
These calls for transparency are because users rarely understand how these complex algorithms function and rarely know how to affect them to better support their own interests and goals.  
Social ``non-technical'' societies, including the European commission, regard  human agency especially, at the audit level as an important requirement for deployed AI systems \cite{hleg2019ethics, fjeld2020principled}. 
For relatable human-human scenarios (e.g. teacher-student, manager-employee), introduction of agency attributes (i.e. perceived self-efficacy) and increasing a sense of agency, has been shown to improve user experience, output productivity and performance \cite{deci2017self}. 
Research shows that the presence of elements such as immediacy behaviours and user agency influence human trust in technological systems positively, even when such systems provide inaccurate results\cite{glikson2020human}.
Still, it is usually an afterthought to design user agency in autonomous situations.

Ridesharing is an example of a laborious activity that is supported by automated AI algorithms for algorithmic management.
There are many stakeholders involved in this activity and they're usually affected by output from these ridesharing AI platforms.
Examples of these stakeholders include drivers and passengers.
Drivers depend on these platforms for work opportunities but they usually do not feel agency over their experiences~\cite{ma2018using}.
This leads to decreased satisfaction and productivity.
To improve these platforms, it's important to initially understand what agency means to drivers and how they navigate current platforms they interact with.
In this preliminary work, using value-sensitive design principles, we interviewed 7 drivers to understand how they handle different ride scenarios and the tools they employ to improve their experiences.



%% file: documentBody/02-RelatedWorks.tex
\section{Background}

\subsection{Agency in Technology}
%

Agency means feeling in control of the outcomes in an environment.
A person is said to feel agency, while utilizing a system, when they are able to perform an action and the reaction the person expects, occurs in the system \cite{richards2016developing}. 
Designing for user agency involves providing novel interactions that are in line with the purpose of the system \cite{gupta2019evaluating}. 
It means the user learns about the system and its inner workings by doing i.e. by interacting and performing actions in the system, thereby improving their perceived self-efficacy. 
Improved self-efficacy signifies that users can take ownership of the outcome and output of a system.

In either critical or normal situations, users should be able to understand, review and challenge AI systems no matter how complex. 
Per the ACM Code of Ethics: ``...all people are stakeholders in computing''~\cite{acm2022}, people and not just developers, should be aware of processes that contribute to the results these systems produce and also, have access to actions that can help modify them. 
They should feel a sense of ownership and responsibility for the system's output.
This combination of factors means that it is important and useful to understand how to give users more agency and control over the complex algorithms and systems that they use.
This will lead to the development of more trustworthy, easy to adopt complex systems. 

\subsection{Value-Sensitive Design}

Value Sensitive Design (VSD) is a theoretical framework that prioritizes ethical human values such as autonomy, bias mitigation, privacy, well-being, trust, accountability etc. in every facet of the technology design process \cite{friedman2002value}. 
It champions the combination of the social and technical aspects when designing a piece of technology so that, the usability and correctness is as important as the ethical human values espoused by the technology \cite{friedman2002value}. 
VSD differs from other approaches that tackle ethics and design like participatory design (pd), computer-supported cooperative work (CSCW), computer ethics and social informatics \cite{friedman2002value}. 
This is because VSD aims to be a part of every step involved in the design process of any kind of technology in any kind of social context, and it enshrines [universal] human values beyond the cooperative and democratic values by pd and CSCW \cite{friedman2002value}. 
According to Friedman et. al \cite{friedman2002value}, the methodology for implementing a value-sensitive design, is an iterative combination of ``conceptual, empirical and technical'' investigations which are defined here: Conceptual investigations involves identifying the direct and indirect stakeholders and investigating the values that are important and can impact them; The empirical aspect uses quantitative and qualitative methods to investigate how people understand or react to the identified values in an interactive environment, how the values relate with usability or correctness issues; In the technical stage, existing technologies are queried to determine if they support or hinder the human values  and/or new technology starts to be developed to support the identified values. 
It is expected that designers iterate through these investigations with no specific order in order to improve their designs \cite{winkler2021twenty}.
A posited extension to the VSD approach includes the inclusion of diverse voices in the selection of values to be considered in a technical design \cite{borning2012next}.

VSD has been applied in the design process of different kinds of system such as a children's online entertainment platform \cite{nouwen2018redefining}, artificial intelligent (AI) systems \cite{liao2019enabling, umbrello2018value, umbrello2019beneficial}, augmented reality \cite{friedman2000new} and autonomous vehicles \cite{thornton2018value}.
Nouwen and Zaman \cite{nouwen2018redefining} provide an example of a way to combine and compare human values from the designer's or corporate's perspective and the end-users.
A novel way to learn about the values of the different stakeholders, especially in futuristic intelligent systems like AI, augmented reality and autonomous vehicles is ``participatory design fictions (PDF)'', where people complete a related fictional story guided by value-centered prompts \cite{liao2019enabling}.
Engaging VSD in the development of AI systems from the initial stages can help to figure the acceptable levels of human values like transparency, control for the stakeholders and also, reduce ``biased or uninformed decisions'' throughout the design process \cite{umbrello2019beneficial, thornton2018value, cummings2006integrating}.

\subsection{Ride-Sharing}

Ridesharing is a type of transportation where a driver shares their private car with a passenger(s) on a trip \cite{cici2015designing}.   
The two main types are static and dynamic \cite{wang2018understanding}. 
In the static type, all the trips are known before-hand while dynamic means the trips come in real-time.
A middle ground between these two types could offer the best possibility for agency and effectiveness.

While examination of how ridesharing platforms affect people, especially drivers, majorly cites lack of agency \cite{ma2018using}, the algorithmic concerns for ride-sharing platforms as observed from recently published research work, are limited to reduced wait-times, travel times, travel cost and maximized profit (through accepting the most number of requests) \cite{luo2022minimizing}. 
Autonomous vehicles (i.e. a scenario where there's zero agency for drivers) are heralded as the future of ridesharing that would solve all these concerns \cite{lokhandwala2018dynamic, singh2019reinforcement}. 
Today, complex algorithms like reinforcement learning, deep learning are routinely applied to the different areas of ridesharing assignment, matching, routing and rating \cite{al2019deeppool} without transparency \cite{lee2015working}.

Some novel algorithmic ridesharing solutions are discussed further.
Wang et. al \cite{wang2018understanding}'s ridesharing framework prioritizes shared trips especially in peak periods.
As new trip requests come in, a scheduling algorithm forms a shared trip for a specified number of people and assigns them to a driver. If a viable shared trip could not be formed in a set amount of time, the rider and driver are assigned a solo trip. 
A driver and possible riders have a fixed travel plan \cite{wang2018understanding} for a set period of time.
Uber has this ride option as uber-x share\footnote{https://www.uber.com/ca/en/ride/uberx-share/}. 
Another mode of shared trips ridesharing is ``multi-hop ridesharing'' \cite{singh2019reinforcement, singh2021distributed} where a passenger's ride to a defined destination is completed in multiple smaller trips (hops). 
This is similar to air travel with connecting flights.
An example of a type of ride-sharing system with existing algorithms that could inform the design of a ride-sharing platform that prioritizes agency is the peer-to-peer (P2P) system. 
In P2P, the drivers have more agency because the riders are matched to the driver's travel plan. 
This requires that the riders register their trips hours until seconds before \cite{tafreshian2020frontiers}. 
In this kind of setting, the drivers are also seen [and treated] as customers \cite{tafreshian2020frontiers}. 
To facilitate some of the prescribed interface solutions, the different kinds of existing peer-to-peer (P2P) matching algorithms \cite{tafreshian2020frontiers} could be useful.
Goel et. al \cite{goel2016privacy} introduce a ridesharing system that prioritizes privacy, an ethical human value related to agency. This privacy option is available for both the driver and rider and privacy means keeping information about identities and locations secure. Their system works by matching drivers, who already provided their location boundaries, and riders based on preferences and ratings after a series of negotiations. The driver and rider are authenticated by a trusted third party (TTP) and they each receive a matching private key. Privacy seems inimical to the ridesharing process but confidentiality and trust is an important feature for vulnerable people such as HIV individuals going for healthcare appointments \cite{f2021examining}.

Existing ride sharing systems like Uber, Lyft are created for profit \cite{gloss2016designing} and so, in cases where stakeholder needs oppose this aim, it is unlikely these needs would be met.
These ridesharing applications delegate management to their algorithmic platforms \cite{ma2018using, lee2015working} but Jhaver et. al \cite{jhaver2018algorithmic}, after observing the same issues including ``algorithmic anxiety'' in another gig-like platform (Air BnB) call for a ``mixed-initiative'' management. 
Many driver workers were attracted to the flexibility nature of ridesharing but the algorithmic management employed in these situations causes loss of agency and various negative issues that impact their wellbeing \cite{ma2022brush, zhang2022algorithmic}.
It is important to examine what alternative algorithmic ridesharing systems that prioritize human needs like control and transparency look like.

%% file: documentBody/03-Method.tex
\section{Method}

Research work \cite{kameswaran2018support, ma2018using, ma2022brush, carlos2021lose, zhang2022algorithmic} on how to improve the well-being of gig workers, including drivers of ridesharing applications, have identified different features and mechanisms that can enhance agency. 
In addition to the information gained from the literature review, it was important to speak directly with drivers. 
In accordance with the value-sensitive design principles, we completed an iteration of a qualitative empirical phase to inform the building of the mobile application in the technological phase. 
We wanted to discover ways they navigated their current platforms, in real world scenarios, to feel agentic and the changes they desired. 
We recruited 7 drivers (all males) through driver groups on the internet. They lived and operated in New York, California, Florida, Pennsylvania and Canada. 
We conducted a semi-structured interview and they were each paid \$20 for an average time of 45 minutes. 
Some of the scenarios  posed to the drivers include: 
\begin{itemize}
    \item You just finished a ride to a downtown area and you received a notification for another ride 5 minutes away. It is 11am on a Saturday. 
    \item You just finished a ride to a downtown area and you received a notification for another ride 15 minutes away. It is 11am on a Saturday.
    \item You are at a busy airport (e.g. JFK) and you received a notification for another ride at a restaurant 10 minutes away. This is your first ride of the day. It is 8pm on a Thursday.
    \item You are at a restaurant and you received a notification for another ride at a busy airport (e.g. JFK) 15 minutes away. This is your first ride of the day. It is 8pm on a Thursday. 
    \item You are at a busy airport (e.g. JFK) and you received a notification for another ride at a restaurant 10 minutes away. You have been riding all day. It is 8pm on a Thursday. 
    \item You are at a restaurant and you received a notification for another ride at a busy airport (e.g. JFK) 15 minutes away. You have been riding all day. It is 8pm on a Thursday. 
\end{itemize}
After each scenario, the drivers were asked if they would accept the rides, they had to provide reasons for their answer choice and then, further questions were asked based on their previous responses. 
Video and audio data were recorded using Zoom. 
The data was transcribed and then, thematically analysed.

%% file: documentBody/04-ResultsDiscussion.tex
\section{Results and Discussion}

Using the combination of findings from the literature review and the qualitative research above, user interface interactions and backend design decisions (which is discussed in the next section) were determined.
We would use the results from this study to illustrate the characteristics of  a driver-centered agentic ridesharing mobile application called Co-opRide.

\subsection{User Interactions}

The following user interface interactions were identified and would be integrated into the Co-opRide mobile application:
\begin{enumerate}
    \item Driver Profile Settings: 
    Aside normal profile information such as name, DOB, licensing and car information, information about the drivers' interest and goals is relevant to making the user experience more agentic. Drivers would be asked their desired weekly or daily earning goal, customer preference (based on ratings they are comfortable with, if the customers give tips, good conversations etc.), number of hours they want to work (part-time vs full-time) and their preferred working hours (e.g. 9-5, nights, etc.). This settings would influence the type of rides and features made available. For example, there would be a ``I m going home'' feature that matches rides along the the driver's home route. Also, the part-time drivers would be more exposed to a ``RideShare'' experience (i.e. getting matched to rides along their route) while full-time drivers would get a ``RideHailing'' experience (i.e. open to rides from a larger circumference area).
    \item Ride Settings and Information: 
    Drivers should be able to set the types of rides they would prefer. This would require settings such as a destination filter, ride lengths and the choice of random ride assignation or queued-up rides. Queued-up rides means another trip is already lined before the driver's current trip is concluded. Drivers have the choice of this setting or a break between rides without penalizing them. 
    In cases where they need to be offered rides outside their preference, reasons should be communicated as well as, a generous incentive.
    When ride notifications are sent to the driver, the system would be transparent about the ride and relevant customer information especially, the specific preference that is violated. Extra ride information such as traffic, remote/urban area, route emergency issues should also, be provided along with more time (>> 45 seconds) so, drivers can make a more informed decision about accepting a ride. Another strategy is providing drivers different trip options at the same time so, they can select the one that best suits them. 
    \item Community Page: According to one of the drivers, it is important to have community to help better navigate current ridesharing platforms but these platforms do not facilitate or accommodate this component in order to avoid the organization of a union. This need for community is echoed in Kameswaran et. al's research work, they discovered that ``drivers often seek and acquire resources from other drivers to help complete their job successfully'' \cite{kameswaran2018support}. Examples of these resources include information about neighbourhoods with more trip options, pickups or destinations to keep away from, words of encouragement and support, etc. The riding application is the best place to facilitate such interaction especially for new drivers that are not aware of platforms such as uberpeople.net, reddit driver groups, therideshareguy.com etc. The community page provided on \textit{Co-op Ride} would be forum-like. Drivers would be able to create topics they are interested in to discuss. There would be location-specific sub forums. If changes were to be made to the application, polls and feedback pages would be introduced to collect drivers' opinions openly.
    \item Complaints Page:
    One of the main issues with algorithmic management is that the ``employees'' are not able to have direct access to their ``employer''. Ridesharing platforms such as Uber \footnote{https://www.uber.com/us/en/drive/driver-app/phone-support/} and Lyft\footnote{https://help.lyft.com/hc/en-us} have started to include in-app and phone support that drivers could use to contact them. Uber also, has ``greenlight hubs'' for in-person meetings but they are mostly available in big cities.
    There needs to be more transparency about how reported issues are handled. \textit{Co-op Ride} would reflect the interactions that allow this such as being able to view the status of one's issue as well as the expected completion time. 
    \item Ratings: A main contention for gig workers including drivers, freelancers, air bnb hosts is the issue of ratings and its opaqueness \cite{ma2018using, carlos2021lose, jhaver2018algorithmic}. Algorithm management companies usually just give a rating without explanation of the factors that contributed to it. Uber recently added a ratings breakdown showing number of people who selected each rating score \cite{singleton_2022}. As the title of the new feature announcement suggests, this is just a ``peek'' and it does not provide useful ``actionable information'' \cite{ma2018using}. Also, the emotional labor usually provided by drivers \cite{ma2022brush, kameswaran2018support} is not distinctly accounted for. There is no clear way to dispute ratings. Some rating changes which will be reflected in the \textit{Co-op Ride} include a more standardized rating scale such that drivers and riders rate the actual factors we believe contribute to a nice driving experience with the option of extra written feedback with prompts like: Describe the most memorable aspect of this trip, etc \cite{ma2018using}. Passengers would be asked to rate the cleanliness, politeness, punctuality, etc. on a satisfaction likert scale (Very dissatisfied, Somewhat dissatisfied, Neither dissatisfied or satisfied, Somewhat satisfied, Very satisfied). Drivers would be notified if certain factors receive continually low scores. Drivers would have the option to dispute a rating if there is verifiable information e.g. time of arrival 
\end{enumerate}

Some of these settings provided to the drivers would be locked for a certain period of time and they would be informed of the time-frame during setup or setting changes. 

\begin{figure}
    \centering
    \begin{subfigure}[b]{0.46\textwidth}
        \centering
        \includegraphics[width=0.6\textwidth]{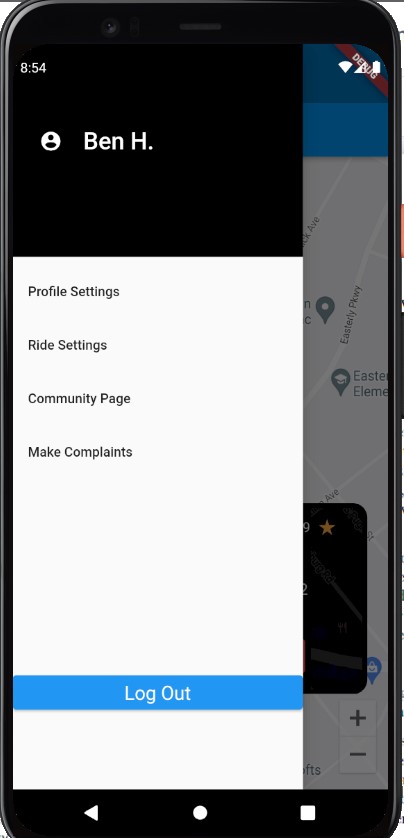}
        \caption{Available App Settings}
        \label{fig:coopsettings}
    \end{subfigure}
    \hfill
    \begin{subfigure}[b]{0.46\textwidth}
        \centering
        \includegraphics[width=0.6\textwidth]{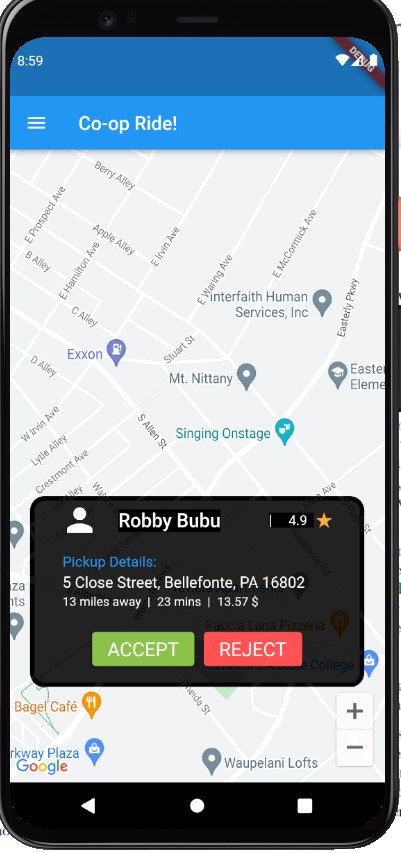}
        \caption{Front Page}
        \label{fig:coopfrontpage}
    \end{subfigure}
    \caption{Co-op Ridesharing Application}
    \label{fig:my_label}
\end{figure}

\subsection{Backend}

Many of the interactions described in the frontend need backend support. 
The database on the backend will store and keep track of each driver, their profile settings, ride preferences, complaints they've submitted, comments they've made in the community page and the ratings they have received. 
To this end, the supported functionalities would be minimal. They are described below:
\begin{enumerate}
    \item Bayesian Model: One of the issues drivers complained about in literature and in our interviews is the random assignment by the ridesharing algorithms. To emulate the factors drivers [try to] consider in their selection of rides, I would use the Bayesian Belief Network (BBN) as the ride matching algorithm.
    A bayesian network is a graphical model made up of nodes, which represent random variables, and edges, which represent the relationship between the variables \cite{brownlee_2019}.
    Bayesian network is the most similar AI technique to the rational way humans reason \cite{wang2019designing}.
    Bayesian networks can be constructed based on expert knowledge, in this case, the experts are the drivers. The aim is to match drivers to rides they are more likely to accept.
    When drivers receive the ride notification, it should include the calculated probability (of the driver accepting the ride) and the top factors that influenced the ride suggestion. Ride incentives starts to be provided when this probability is less than 0.6 (an arbitrary number which can be voted on in the community page).
    The random variables and the relationships identified as a result of the interview process are shown in Figure \ref{fig:bayes}.
    The system would use the driver's profile settings to dictate the expert knowledge the bayesian network starts with. As observed conditions (drivers clicking accept or cancel on simulated rides) are recorded the network would learn to fit the driver.
    
    \begin{figure}[h]
    \centering
    \includegraphics[width=\columnwidth]{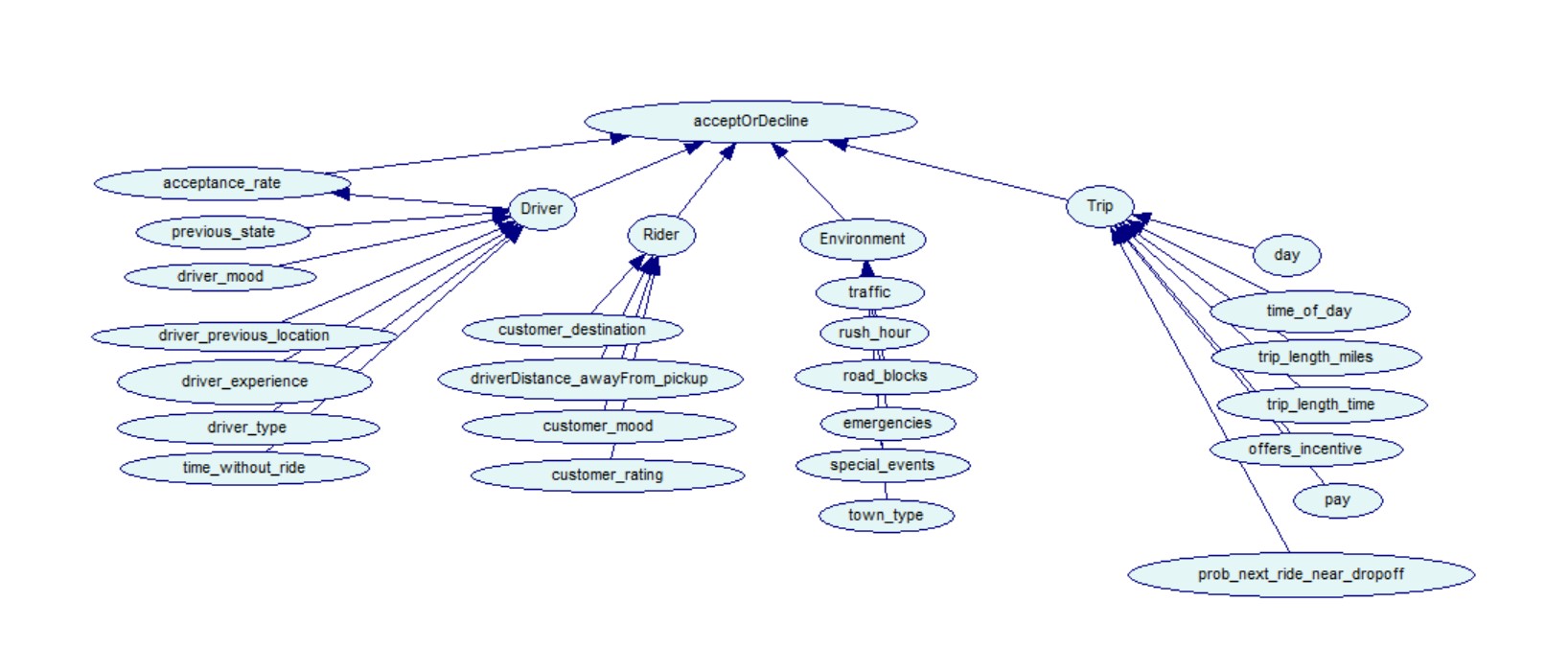}
    \caption{Bayesian Network showing random variables that determine driver's acceptance of ride}
    \label{fig:bayes}
\end{figure}
    
    \item Earning Model: 
    Transparency is a major part of agency. Current ridesharing systems e.g. Uber \footnote{https://www.uber.com/us/en/drive/how-much-drivers-make/}, Lyft \footnote{https://www.lyft.com/driver/pay\#pay} do not provide enough information about their price model. They mention the factors they consider such as a base fare, trip length and possible rider wait time. The gas and maintenance are said to be the driver's responsibility.
    This is unfair and opaque. A more rounded pricing model, which will be utilized in \textit{Co-op Ride}, was introduced by Southern et. al \cite{southern2017understanding}. This pricing model outputs ``Total Cost of Ownership'' (TCO) per trip using fixed costs such as car depreciation rate, insurance cost, taxes and per trip costs such as fuel cost and maintenance \cite{southern2017understanding}.
    
    Drivers have called the challenges and promotions provided to earn bonuses in current ridesharing systems unfair, because it usually forces drivers to work longer hours \cite{ma2018using}. A standard yearly or monthly bonus could be provided instead, to all drivers based on the number of hours they worked.
\end{enumerate}

%% file: documentBody/05-Conclusion.tex
\section{Conclusion and Future Work}

In this preliminary work, we examined what agency means to drivers in a ridesharing application.
The results show that drivers want control over their work-time, the passengers they pick, transparency (pay, algorithm, etc.), they type of rides they're offered, a page to congregate with other drivers, a more effective way to seek redress and so on.
Future work would study and expand on the effects, of enabling these control functions, on the driver, passengers and platform owners.